\documentclass[10pt,letterpaper]{article}
\usepackage[top=0.85in,left=2.75in,footskip=0.75in]{geometry}

\usepackage{amsmath,amssymb}

\usepackage{changepage}

\usepackage{textcomp,marvosym}

\usepackage{cite}

\usepackage{nameref,hyperref}

\usepackage[right]{lineno}

\usepackage[nopatch=eqnum]{microtype}
\DisableLigatures[f]{encoding = *, family = * }

\usepackage[table]{xcolor}

\usepackage{array}

\newcolumntype{+}{!{\vrule width 2pt}}

\newlength\savedwidth

\raggedright
\usepackage[aboveskip=1pt,labelfont=bf,labelsep=period,justification=raggedright,singlelinecheck=off]{caption}

\makeatletter
\renewcommand{\@biblabel}[1]{\quad#1.}
\makeatother

\usepackage{lastpage,fancyhdr,graphicx}
\usepackage{epstopdf}
\fancyheadoffset[L]{2.25in}
\fancyfootoffset[L]{2.25in}
\usepackage{multirow}
\usepackage{booktabs}

\def\Si{\mathsf S}
\def\Ei{\mathsf E}
\def\Ii{\mathsf I}

\def\Ni{\mathsf N}
\def\R{\mathcal{R}}
\def\Ro{\R_{0}}
\def\dd{\mathrm{d}}

\begin{document}
\vspace*{0.2in}

\begin{flushleft}
{\Large
% \textbf{Does choice of susceptibility distribution in heterogeneous epidemic models matter in forecast?}
\textbf{An exploration into how susceptibility distribution misspecifications impact epidemic forecasting}
}
\newline
\\
Ibrahim Mohammed\textsuperscript{1,2},
Chris Robertson\textsuperscript{1,3},
M. Gabriela M. Gomes\textsuperscript{1,4*}
\\
\bigskip
\textbf{1} Department of Mathematics and Statistics, University of Strathclyde, Glasgow, G1 1XH, UK
\\
\textbf{2} Department of Mathematical Sciences, Abubakar Tafawa Balewa University, Bauchi, Nigeria
\\
\textbf{3} Public Health Scotland, Glasgow, G2 6QE, UK
\\
\textbf{4} Centre for Mathematics and Applications (NOVA MATH), NOVA School of Science and Technology, Caparica, 2829-516, Portugal
\\
\bigskip

% Use the asterisk to denote corresponding authorship and provide email address in note below.
* gabriela.gomes@strath.ac.uk

\end{flushleft}

\section*{Abstract}
Heterogeneous susceptibility models for epidemic dynamics preferentially assume that individual susceptibility follows a gamma distribution, which permits analytical reduction to a low-dimensional system. However, the true empirical distributional form in any given population is unknown. Here we investigate the consequences of misspecifying the susceptibility distribution by comparing gamma and lognormal specifications in a Susceptible-Exposed-Infectious-Removed (SEIR) framework. When both distributions are matched on mean and coefficient of variation ($\nu$), we find that their epidemic trajectories diverge once heterogeneity is moderate or high ($\nu \gtrsim 1$), with the lognormal producing a later, larger peak and a greater final size. We then assess the impact of distributional misspecification on statistical inference. Using synthetic datasets, we fit correctly specified and misspecified models by maximum likelihood. In a default scenario, where inference is based on simulated data for a single epidemic, both models can reproduce the data by compensating through correlated shifts in heterogeneity and intervention parameters. When inference is based on two simulated epidemics, however, this compensation may be reduced by known constraints of how parameters are related across epidemics. In these cases, the correctly specified model recovers all parameters accurately, while the misspecified model tends to give biased estimates. These inference biases propagate into forecasts, but predictions remain relatively accurate when compared to homogeneous models which more than double peak incidences in scenarios where $\nu \approx 1$, for instance. We conclude that deviations resulting from the susceptibility distribution misspecifications assessed here are minor and encourage the adoption of heterogeneous models in future epidemic forecasting.

% Please keep the Author Summary between 150 and 200 words. Use first person.
% PLOS ONE authors: please skip this step.
%
\section*{Author summary}
We investigated whether the mathematical form assumed for how susceptibility varies across individuals matters for epidemic modelling. Many models assume a gamma distribution for convenience, but in practice the true shape is unknown. We compared gamma and lognormal distributions in a standard SEIR epidemic model and found that, when heterogeneity is moderate or high, these two choices produce meaningfully different epidemic trajectories when their mean and variance are identical. Fitting the wrong distributional family to simulated data reveals that a single epidemic wave cannot distinguish between the two, because parameter correlations absorb the structural mismatch. However, jointly fitting two concurrent epidemics may reduce this compensation and expose the misspecification through collapsed confidence interval coverage. Despite this, the forecast cost of using the wrong heterogeneous family remains much smaller than the cost of ignoring heterogeneity altogether. Our results support using heterogeneous models as a default and treating distributional choice as a secondary, though not negligible, modelling decision.

\clearpage
\newgeometry{top=0.85in,left=1in,right=1in,footskip=0.75in}
% \linenumbers

% Use "Eq" instead of "Equation" for equation citations.
\section{Introduction}
 
In a recent paper \cite{Mohammed2025}, we investigated the identifiability of heterogeneity in susceptibility to infection and impact of non-pharmaceutical interventions (NPIs) using SEIR models with gamma-distributed susceptibility, as analysed in \cite{Antonio2022Reinfection} and applied in \cite{Gomes2022IndividualThreshold}. That study showed that when data are generated under heterogeneous susceptibility, a homogeneous model misattributes the effect of heterogeneity to NPIs, and that fitting two concurrent epidemics with shared parameters alleviates the identifiability issues.
 
The choice of gamma distributions for susceptibility is also seen in models developed to study the efficacy of vaccines \cite{Halloran1996, Langwig2017} and to analyse dose-infectivity curves generated from controlled experiments \cite{Dwyer1997, BenAmi2010, King2018, Miura2025}. Mathematically, it is an attractive distribution \cite{Novozhilov2008OnPopulation, Margheri2017, Antonio2022Reinfection}. It has also been shown to act as an ``eigen-distribution'' under selective depletion \cite{Rose2021HeterogeneityModels}, preserving its functional form as highly susceptible individuals are removed from the population. This allows the explicit infinite SEIR system to be reduced to a closed-form low-dimensional ODE system \cite{Novozhilov2008OnPopulation, Antonio2022Reinfection}, making inference computationally tractable.
 
Despite its computational benefits, this closed-form is inherently limited to studying closed epidemics over short periods of time. Extending the model to capture longer-term phenomena such as vital dynamics involving birth and death rates or waning acquired immunity conflicts with the conditions required for the low-dimensional closed-form. The system can still be analysed numerically in these cases, but this requires discretising the continuous susceptibility distribution into a finite number of interacting groups. Essentially, this discretisation scheme does more than just pave the way for demographic turnover; it removes the structural reliance on the gamma family, making it possible to study the model using arbitrary susceptibility distributions, as demonstrated in our misspecification analysis.
 
This flexibility is important because the true empirical form of the susceptibility distribution in any real population is unknown. Several alternatives have been proposed, including the lognormal distribution \cite{McGilchristAisbett1991}, the beta distribution as a bounded alternative \cite{Furumoto1967, Novozhilov2008OnPopulation, Pessoa2014, Gomes2014, Margheri2015, Langwig2017, Tekeli2025}, and the inverse Gaussian distribution from frailty theory \cite{Hougaard1984}. Even when two candidate distributions are calibrated to share the same mean and coefficient of variation ($\nu$), they differ in their higher-order moments (skewness and kurtosis), which result in different epidemic dynamics.
 
More broadly, the consequences of structural misspecification in infectious disease models have received growing attention across several modelling traditions.  Misspecification within pairwise accelerated failure time (AFT) models for infectious disease transmission was studied in \cite{sharker2024pairwise}. The authors showed that incorrect assumptions about the contact interval distribution can bias mechanistic inference even when overall epidemic patterns are reproduced accurately. This work demonstrated that regression coefficients for susceptibility remain robust under moderate misspecification provided the fitted model retains sufficient parametric flexibility (e.g.\ a Weibull model applied to log-logistic data), while quantities directly tied to the underlying transmission mechanism, such as hazard structure, transmission timing, and secondary attack risks, are sensitive to the assumed distributional family. In a complementary direction, misspecification was also seen to arise from the statistical assumptions imposed during inference \cite{Gao2023Misspecification} rather than from the biological model itself. In Bayesian phylodynamic models used to infer the spatial history of disease outbreaks, the default prior distributions on dispersal rates and route indicators encode strong and biologically unrealistic assumptions about the geographic process. Because these models contain many parameters estimated from minimal data (a single geographic observation per sampled pathogen), the posterior is inherently prior-sensitive, and the misspecified priors qualitatively distort key epidemiological conclusions, including the ancestral origin of outbreaks, the relative importance of dispersal routes, and the number of inter-area transmission events.
 
These studies reveal a recurring pattern: misspecified models can reproduce observed data by compensating through other parameters, creating a false sense of model adequacy. In \cite{sharker2024pairwise}, the wrong contact interval distribution is absorbed by shifts in rate parameters, preserving some regression effects while distorting others. In \cite{Gao2023Misspecification}, overly informative priors constrain the posterior in ways that the sparse data cannot override, producing confident but misleading geographic histories. The issue is especially relevant in heterogeneous epidemic models because selective depletion dynamically reshapes the effective susceptible population over time. Consequently, two susceptibility distributions with identical first and second moments may still result in different epidemic trajectories through differences in their tail behaviour and higher-order structure. This raises the possibility that epidemic data from a single outbreak may be insufficient to reliably distinguish between competing heterogeneity assumptions, particularly when other mechanisms such as NPIs can compensate for misspecification during inference.
 
Here we address two practical questions. First, when heterogeneity is moderate to high (say, $\nu gtrsim 1$), how are epidemic trajectories affected by higher moments of the susceptibility distribution? Second, if the wrong distributional family is assumed during inference, what are the consequences for parameter estimation and forecasting -- and does the answer change when multiple epidemics with shared parameters are fitted jointly?
 
We address these questions using the same SEIR model as in \cite{Antonio2022Reinfection, Gomes2022IndividualThreshold, Mohammed2025}, extended to accommodate arbitrary susceptibility distributions through a discretisation scheme (described in \nameref{S1_Appendix}). We compare gamma and lognormal distributions throughout. Our main findings are as follows. At the trajectory level, lognormal susceptibility produces a higher and later incidence peak and a larger final epidemic size than gamma susceptibility, for the same mean and coefficient of variation. At the inference level, single-epidemic fitting can mask misspecification because parameter correlations allow compensating shifts in the heterogeneity and NPI estimates. Two-epidemic fitting reduces this flexibility by narrowing the confidence intervals and making the misspecified model appear biased. When compared to homogeneous models, however, the deviations due to the distribution misspecifications assessed here are minor, encouraging the adoption of heterogeneous models in future epidemic forecasting.

\section{Mathematical models}

\subsection{SEIR model with heterogeneous susceptibility}\label{sec:model}

We use the same model as in \cite{Antonio2022Reinfection, Gomes2022IndividualThreshold, Mohammed2025}. Let $q(x)$ be a probability density function representing the distribution of individual susceptibility $x$ across the population, with mean 1 and coefficient of variation
\begin{equation}\label{eq:cv}
    \nu = \sqrt{\int (x-1)^2 q(x)\,\dd x}.
\end{equation}
The heterogeneous SEIR model is written in terms of differential equations as
\begin{equation}\label{eq:het_system}
\begin{split}
 \frac{\dd S(x)}{\dd t}&= - x\, c(t)\, \beta\, (\rho\, \Ei+\Ii)\, \frac{S(x)}{\Ni}, \\
 \frac{\dd\Ei}{\dd t}&=  x\, c(t)\, \beta\, (\rho\, \Ei+\Ii)\, \frac{S(x)}{\Ni} - \delta\, \Ei, \\
 \frac{\dd\Ii}{\dd t}&= \delta\, \Ei-\gamma\, \Ii,
\end{split}
\end{equation}
where $S(x)$ represents the density of susceptible individuals as a function of the susceptibility trait $x$, and $\Si = \int S(x)\,\dd x$. The main parameters are the average effective contact rate $\beta$, the rate of progression from $\Ei$ to $\Ii$ (assumed $\delta=1/5.5$ per day), the rate of removal from $\Ii$ (assumed $\gamma=1/4$ per day) and the reduced infectiousness while in $\Ei$ (assumed $\rho = 0.5$). The basic reproduction number is
\begin{equation}\label{eq:R0}
\Ro= \beta\left(\frac{\rho}{\delta}+\frac{1}{\gamma}\right).
\end{equation}

When $q(x)$ is a gamma distribution, system~\eqref{eq:het_system} reduces exactly to the low-dimensional system of ordinary differential equations (ODEs) written as
\begin{equation}\label{eq:reduced}
\begin{split}
 \frac{\dd \Si}{\dd t}&= -c(t)\, \beta\, (\rho\, \Ei+\Ii)\, \left(\frac{\Si}{\Ni} \right)^{1+\nu^2}, \\
 \frac{\dd \Ei}{\dd t}&= c(t)\, \beta\, (\rho\, \Ei+\Ii)\, \left(\frac{\Si}{\Ni} \right)^{1+\nu^2} - \delta\, \Ei, \\
\frac{\dd \Ii}{\dd t}&= \delta\, \Ei-\gamma\, \Ii,
\end{split}
\end{equation}
as shown in \cite{Novozhilov2008OnPopulation, Antonio2022Reinfection}. This reduction is specific to the gamma family. For other distributions, such as the lognormal, one must work with the explicit system~\eqref{eq:het_system}, which can be approximated by a finite system of ODEs by discretising $q(x)$ into a finite number of susceptibility classes.

\subsection{Non-pharmaceutical interventions}

NPIs are modelled using a time-dependent factor $c(t)$ as in \cite{Gomes2022IndividualThreshold, Mohammed2025}:
\begin{equation}\label{eq:npi}
c(t) =
\begin{cases}
  1, & \text{if } 0 < t \leq t_0, \\
  \displaystyle 1 - (1-c^{\ast})\frac{t-t_0}{t_1-t_0}, & \text{if } t_0 < t \leq t_1, \\
  c^{\ast}, &  \text{if  } t_1 < t,
\end{cases}
\end{equation}
where $0\leq c^{\ast}\leq 1$, $t_0$ is the time when contact rates begin to decrease and $t_1$ marks the beginning of maximal containment such as lockdown. A description of all model parameters is provided in Table~\ref{tab:params}.

% Place tables after the first paragraph in which they are cited.
\begin{table}[!ht]
\centering
\caption{
\textbf{Model parameters motivated by empirical estimates from the COVID-19 pandemic.}
In particular, $\Ro \approx 3.0$ reflects early transmission estimates in the UK, while the coefficient of variation $\nu = 1.414$ is consistent with reported levels of heterogeneity in susceptibility. Other epidemiological parameters are taken from \cite{Gomes2022IndividualThreshold}.}
\label{tab:params}
\begin{tabular}{|l|l|l|l|}
\hline
\textbf{Parameter} & \textbf{Description} & \textbf{Value} & \textbf{Source / notes} \\
\hline
$N$ & Total population & $100{,}000$ & Assumed \\
\hline
$\delta$ & Rate $\Ei\!\to\!\Ii$ & $1/5.5$ per day & \cite{Gomes2022IndividualThreshold}\\
\hline
$\gamma$ & Recovery rate from $\Ii$ & $1/4$ per day & \cite{Gomes2022IndividualThreshold} \\
\hline
$\rho$ & Relative infectiousness in $\Ei$ & $0.5$ & \cite{Gomes2022IndividualThreshold}\\
\hline
$\Ro$ & Basic reproduction number & 3.0 & Fixed \\
\hline
$\nu$ & Coefficient of variation & $\sqrt{2}\approx 1.414$ & Fixed \\
\hline
$c_1$ & Sustained contact multiplier & 0.3 & Fixed \\
\hline
$t_0$ & Behavioural change onset time & $15$ days & Fixed \\
\hline
$t_1$ & Time to beginning of lockdown & $20$ days & Fixed \\
\hline
\end{tabular}
\end{table}

\subsection{Discretisation of the susceptibility distribution}\label{sec:discretisation}

When the susceptibility distribution is not gamma, the reduced system~\eqref{eq:reduced} no longer applies and one must solve the explicit system~\eqref{eq:het_system}. We discretise the continuous distribution $q(x)$ into $k$ classes with weights $\{q_i\}$ and representative susceptibilities $\{x_i\}$, converting system~\eqref{eq:het_system} into $k+2$ ODEs.

The discretisation uses a three-stage transformation: (i) partition the CDF to obtain group probabilities $\{q_i\}$ (defined in supporting information) ;(ii) assign each group a representative $x_i$ as the conditional expectation within that interval; (iii) calibrate the representatives to match the target mean ($\mu_\star = 1$) and target variance ($\sigma_\star^2 = \nu^2$) using a log-affine transformation $x_i' = \exp(A + B \log x_i)$, where $A$ and $B$ are determined by solving two moment constraints. This scheme preserves positivity and ordering. Technical details are provided in \nameref{S1_Appendix} and \cite{ibrahimThesis}.

For gamma targets, we verified that the discretised system~\eqref{eq:het_system} reproduces the trajectories of the closed form system~\eqref{eq:reduced} to high accuracy once $k \geq 20$, with errors declining rapidly with $k$ (see \nameref{S1_Appendix}). Cross-family comparisons between gamma and lognormal require finer resolution ($k \geq 40$) to avoid underestimating the structural differences between distribution families.

\subsection{Gamma vs lognormal: distributional differences}\label{sec:distributions}

Both distributions are parameterised to have mean 1 and a prescribed coefficient of variation $\nu$. For the gamma distribution this corresponds to shape parameter $\alpha = 1/\nu^2$ and rate $\beta = \alpha$. For the lognormal distribution we require parameters $\sigma^2 = \log(1+\nu^2)$ and $\mu = -\sigma^2/2$, where $\mu$ is the logarithm of location and $\sigma$ is the logarithm of scale.

\begin{figure}[!h]
\centering
\includegraphics[width=0.95\linewidth]{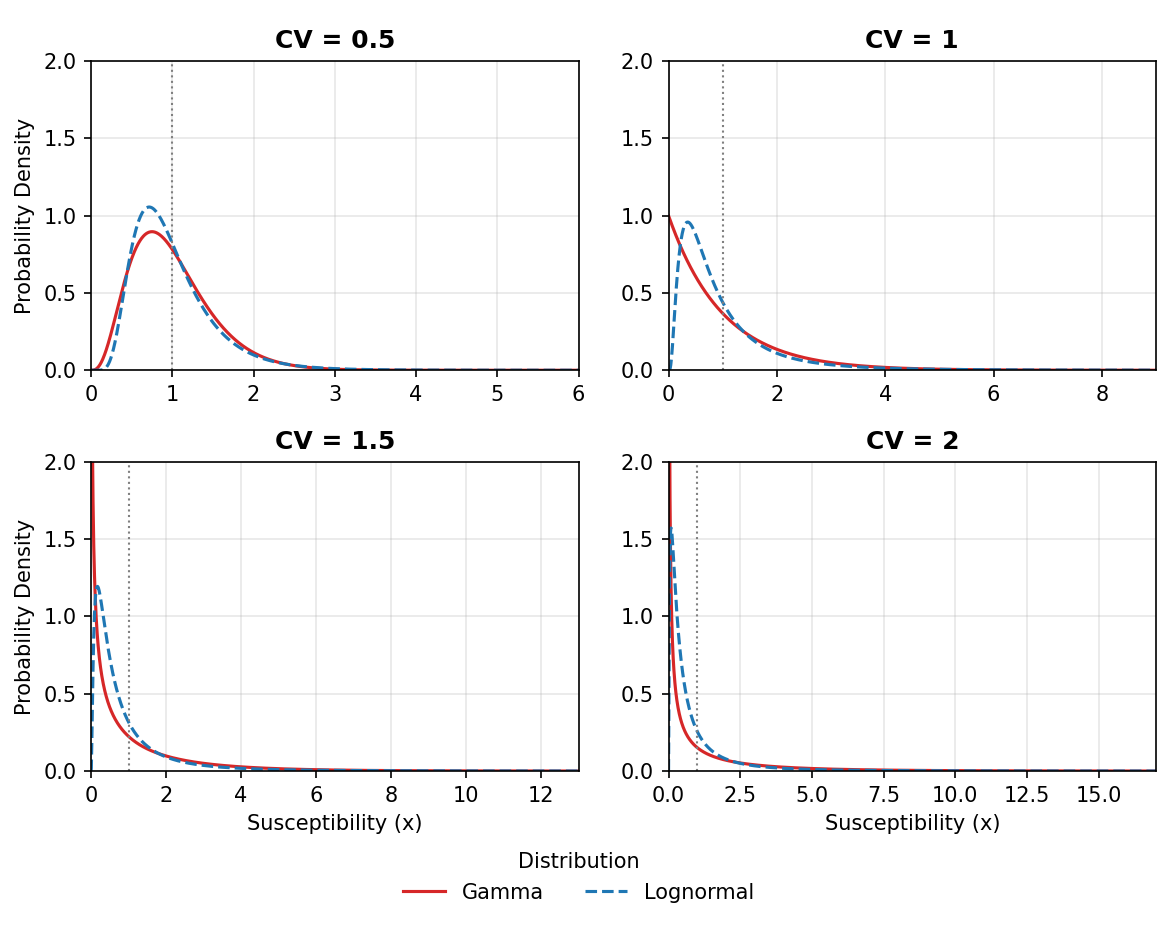}
\caption{
\textbf{Gamma and lognormal probability density functions with identical means (1) and coefficients of variation ($\nu$).}
The lognormal exhibits heavier right tails, especially at higher $\nu$.}
\label{fig:pdf_comparison}
\end{figure}

Although the first two moments are matched, the higher-order moments differ. The lognormal has heavier tails than the gamma at the same $\nu$, with higher skewness and kurtosis, as illustrated in Fig~\ref{fig:pdf_comparison}.

% Place figure captions after the first paragraph in which they are cited.

\begin{figure}[!h]
\centering
\includegraphics[width=0.95\linewidth]{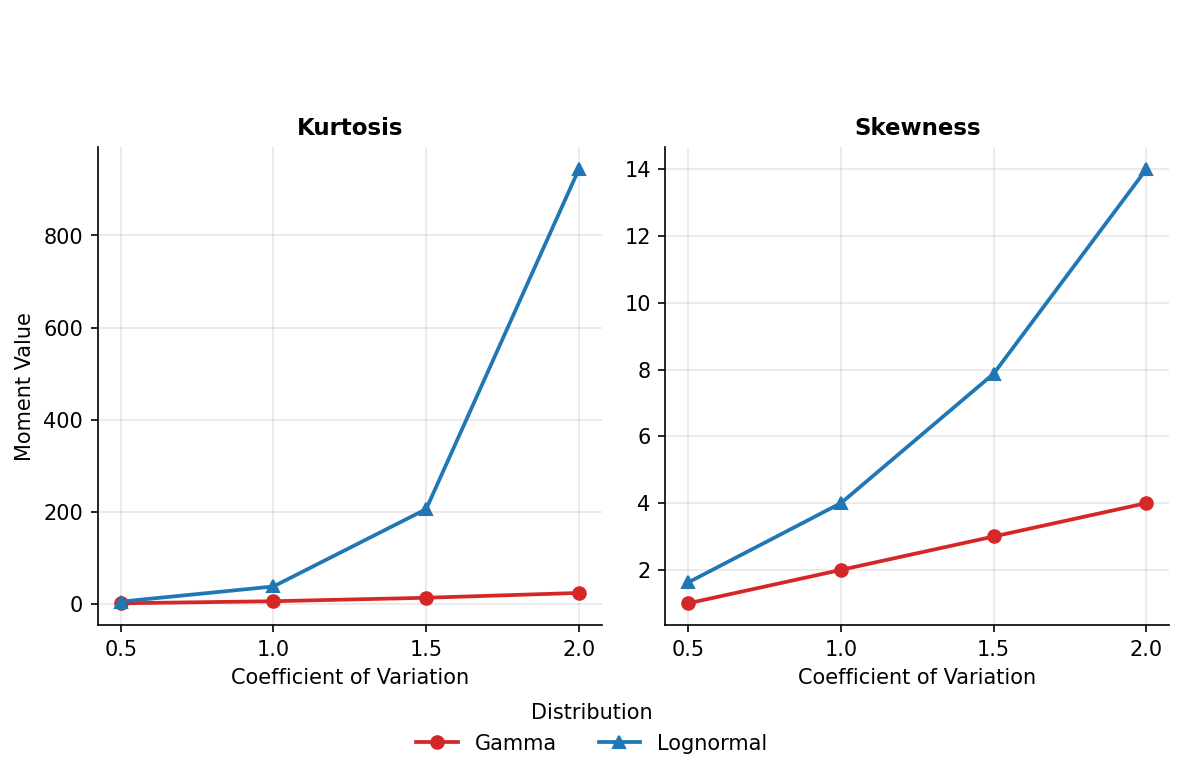}
\caption{
\textbf{Higher-order moment differences between gamma and lognormal distributions.}
With both distributions having identical means and variances by construction, their skewness and kurtosis diverge markedly with increasing $\nu$. The lognormal shows exponentially increasing skewness and kurtosis, indicating different tail behaviours.}
\label{fig:moments_comparison}
\end{figure}

Fig~\ref{fig:moments_comparison} quantifies the higher-order moment differences. With both distributions having identical means and variances by construction, their third and fourth moments (skewness and kurtosis) differ markedly. The lognormal distribution shows exponentially increasing skewness and kurtosis with the coefficient of variation, indicating different tail behaviours that will drive diverging epidemic trajectories as shown in the next section.

\section{Exploration of how epidemic trajectories diverge} \label{sec:results_traj}

We first show whether the shape of the susceptibility
distribution (holding the first two moments equal) produces
systematic differences in deterministic epidemic trajectories.
We compare gamma and lognormal versions of
system~\eqref{eq:het_system} with no NPIs ($c^\ast = 1$) and
$\Ro = 3$, varying $\nu$ across $\{0.5,\,1.0,\,1.414,\,2.0\}$ and
discretisation resolution $k$ across $\{10,\,20,\,40,\,60,\,100\}$.

We write $C_\bullet(t)$ for total cumulative incidence under
distribution $\bullet\in\{\Gamma,\mathrm{LN}\}$ and define daily
incidence as $y_\bullet(t) = C_\bullet(t) - C_\bullet(t-1)$, with
$y_\bullet(0)=0$. We use three summary metrics quantify the cross-family differences:

\begin{enumerate}
  \item \textbf{Relative peak magnitude.}
        The percentage difference in peak daily incidence,
        \[
          \Delta y_{\mathrm{peak}}^{(\%)}
          = 100\;\frac{\max_t y_{\mathrm{LN}}(t)
                      - \max_t y_{\Gamma}(t)}
                      {\max_t y_{\Gamma}(t)}.
        \]

  \item \textbf{Peak timing shift.}
        The difference in the day of peak incidence,
        \[
          \Delta t_{\mathrm{peak}}
          = t_{\mathrm{LN}}^{\max} - t_{\Gamma}^{\max},
          \qquad
          t_\bullet^{\max} = \arg\max_t\, y_\bullet(t).
        \]

  \item \textbf{Relative final size.}
        The percentage difference in cumulative incidence at the
        simulation horizon $T$,
        \[
          \Delta C_T^{(\%)}
          = 100\;\frac{C_{\mathrm{LN}}(T) - C_{\Gamma}(T)}
                      {C_{\Gamma}(T)}.
        \]
\end{enumerate}

\noindent
Positive values in all three metrics would indicate that the lognormal model produces a larger, later epidemic than the gamma model at the same $\nu$.

\subsection{Pattern of divergence}
\label{sec:divergence}

\begin{figure}[!h]
\centering
\includegraphics[width=0.95\linewidth]{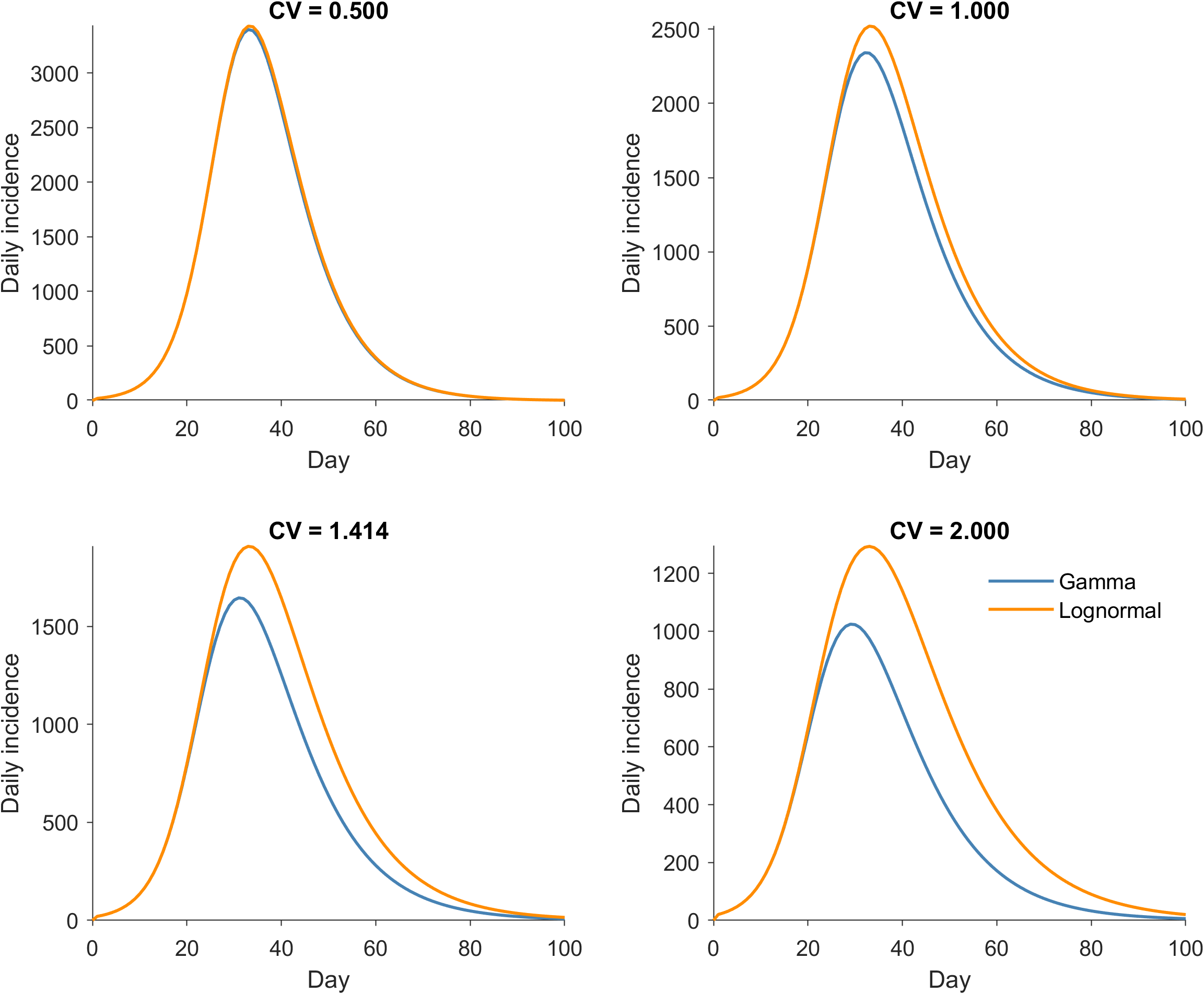}
\caption{
\textbf{Daily incidence trajectories from discretised gamma (blue) and lognormal (red dashed) heterogeneous SEIR models, using discretisations into $k=60$ bins.}
Parameters as in Table~\ref{tab:params} with no NPIs ($c(t)\equiv 1$). At low $\nu$ the curves are similar; at high $\nu$ the lognormal produces a larger, later peak. }
\label{fig:trajectory}
\end{figure}

When $\nu$ is low ($\nu \leq 0.5$), the two distributions produce nearly identical epidemic curves. As heterogeneity increases, a clear pattern emerges (Fig~\ref{fig:trajectory}):
the lognormal model produces a higher peak incidence than the gamma model, peaks later, and results in a larger final size. These differences grow with $\nu$.

Mechanistically, this difference is driven by the tail structure. The gamma distribution depletes its high-susceptibility core rapidly through selective removal, causing the effective reproduction number to drop early. The lognormal distribution, with its heavier right tail, contains more highly-susceptible individuals who sustain transmission chains longer, preventing the epidemic from slowing down as quickly. 

\subsection{Convergence of results with resolution of discretisation}

\begin{figure}[!h]
\centering
\includegraphics[width=\linewidth]{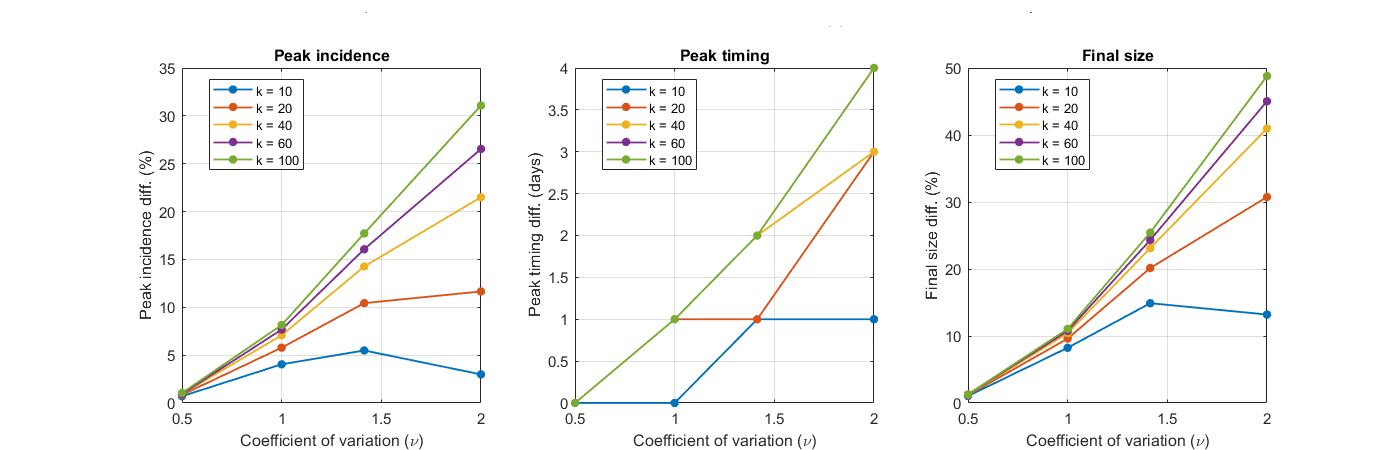}
\caption{
\textbf{Percentage difference (lognormal relative to gamma) in peak incidence (left), peak timing (centre) and epidemic final size (right), across $\nu$ and discretisation resolution $k$.}
The lognormal consistently produces larger outcomes. Differences are negligible at low $\nu$ but substantial at high $\nu$ once $k \gtrsim 40$.}
\label{fig:summary_metrics}
\end{figure}

Fig~\ref{fig:summary_metrics} summarises the percentage differences in peak incidence, peak timing, and final size across $\nu$ and $k$. At $k = 10$, the coarse grid underestimates the cross-family difference, especially at high $\nu$, because the lognormal tail is compressed into too few bins. By $k \geq 40$, the metrics appear to converge and grow approximately linearly with $\nu$. By $\nu = 2$, the lognormal peak incidence is approximately 30\% higher and the final size approximately 50\% larger than that given by the gamma model.

A practical implication is that while $k \approx 20$ suffices for within-family trajectory matching (i.e., verifying that the discretised gamma reproduces the reduced gamma model), cross-family comparisons require $k \gtrsim 40$ to avoid masking the structural differences.

% \subsection{Trajectory comparison with NPIs}\label{sec:traj_npi}

% Figs~\ref{fig:trajectory} and~\ref{fig:summary_metrics} characterised the cross-family divergence for an unmitigated epidemic ($c(t) \equiv 1$). Fig~\ref{fig:trajectory_npi} shows that the same qualitative pattern persists when NPIs are active and extends into the forecast period once interventions are lifted. Fits from the gamma and lognormal heterogeneous models are visually indistinguishable over the training window (days 1--100); after NPIs are removed at day~100, the lognormal fit projects a larger and later second wave than the gamma fit, consistent with the heavier-tail mechanism identified above. This motivates the inference experiments that follow: fitting the wrong family does not only mis-estimate heterogeneity, it carries directional consequences for any projection beyond the training window.

% \begin{figure}[!h]
% \centering
% \includegraphics[width=0.85\linewidth]{Single_ep/prediction_trajectories_gamma_truth.pdf}
% \caption{
% \textbf{Fitted training window (days 1--100) and forecast (days 101--250) trajectories at baseline heterogeneity ($\Ro = 3$, $\nu = 1.414$, $c^{\ast} = 0.3$).}
% Solid lines are mean fitted/forecast trajectories; shaded bands are 95\% confidence envelopes across replicated fits. After NPIs are lifted at day~100 (vertical dashed line), the lognormal fit projects a larger and later second wave than the gamma fit.}
% \label{fig:trajectory_npi}
% \end{figure}

\section{Impact of misspecification on inference} \label{sec:results_inf}

Section \ref{sec:divergence} demonstrated that gamma and Log-normal susceptibility specifications produce measurably different trajectories at moderate to high $\nu$. We now quantify how these structural differences translate into statistical inference when the susceptibility distribution family is misspecified.

\subsection{Simulation and observation model} \label{sec:datagen}

For a given truth specification (gamma or lognormal), we integrate system~\eqref{eq:het_system} with NPIs to obtain the mean daily incidence. Let $\Ei(t)$ denote the number of exposed individuals at day $t$. Then the model-implied mean daily incidence is
\begin{equation}\label{eq:incidence}
    x_t(\boldsymbol{\theta}) \;=\; \delta\ \Ei(t),
\end{equation}
where $\delta$ is the rate of progression from $\Ei$ to $\Ii$ (mean latent period $1/\delta$). The observed incidence is generated via a Poisson sampling model,
\begin{equation}\label{eq:obs_model}
    y_t \sim \mathrm{Poisson}\!\big(x_t(\boldsymbol{\theta})\big), \qquad t = 1, \dots, T.
\end{equation}
We use $\Ro = 3$, $\nu = 1.414$, $c^{\ast} = 0.3$, $t_0 = 15$ (Table~\ref{tab:params}) and $T=100$~days, and repeat this procedure independently across 200 replicated datasets for each truth scenario.

\subsection{Inference target and likelihood} \label{sec:likelihood}

Inference is performed by maximum likelihood estimation. For both discretised gamma and lognormal susceptibility models, we estimate the parameter vector
\begin{equation}\label{eq:theta}
    \boldsymbol{\theta} := (\Ro,\, \nu,\, t_0,\, c^{\ast}).
\end{equation}
Assuming the observations $y_t$ are conditionally independent given the model states $x_t(\boldsymbol{\theta})$, the likelihood function is
\begin{equation}\label{eq:likelihood}
\mathcal{L}(\boldsymbol{\theta}) = \prod_{t=1}^{T} f\!\left(y_t \mid x_t(\boldsymbol{\theta})\right),
\end{equation}
where $f(\cdot)$ is the Poisson probability mass function and $x_t(\boldsymbol{\theta})$ is the model-implied daily incidence from Eq~\eqref{eq:incidence}. The log-likelihood is
\begin{equation}\label{eq:loglik}
    \ell(\boldsymbol{\theta}) \;=\;
    \sum_{t=1}^{T} \left[
        y_t \log x_t(\boldsymbol{\theta}) - x_t(\boldsymbol{\theta})
    \right],
\end{equation}
dropping terms not depending on $\boldsymbol{\theta}$.

This likelihood is then evaluated over epidemic trajectories as in \cite{Mohammed2025}. Approximate 95\% confidence intervals are obtained from the observed Fisher information (inverse Hessian) at the maximum likelihood estimation (MLE). Performance is summarised by the mean and standard deviation (SD) of MLEs across 200 datasets, mean 95\% CI width, and empirical coverage. We adopt a $2 \times 2$ truth--fit design: data generated from either gamma or lognormal distributions, fitted with each family.

\subsection{Single-epidemic inference} \label{sec:single_misspec}

Here we conduct our analyses in the standard scenario in which only one epidemic is available for inference. The main analysis uses the lognormal-distributed susceptibility model to generate synthetic data and fits three candidate models: a correctly specified lognormal model, a misspecified gamma model, and a homogeneous reference model with $\nu = 0$. The gamma-truth scenario, in which the directional roles of the two heterogeneous families are reversed, is reported in \nameref{S1_Appendix}.
Table~\ref{tab:single_logn} summarises the three fits across 200 replicated datasets at $\nu_{\text{true}} = 1.414$.

\begin{table}[!ht]
\centering
\caption{
\textbf{Single-epidemic inference for lognormal-generated data ($\nu_{\text{true}} = 1.414$).}
Median (SD), mean 95\% CI width, and empirical coverage across 200 replicates. Coverage is the percentage of replicates whose 95\% confidence interval contains the true parameter value. Convergence (positive-definite Hessian): gamma 183/200, lognormal 176/200, homogeneous 200/200. The homogeneous fit fixes $\nu = 0$.}
\label{tab:single_logn}
\begin{tabular}{|l|l|l|l|l|}
\hline
\textbf{Model} & \textbf{Parameter} & \textbf{Median (SD)} & \textbf{CI width} & \textbf{Coverage (\%)} \\
\hline
\multirow{4}{*}{Lognormal (correct)}
  & $\Ro$       & 2.996 (0.034) & 0.116 & 84.7 \\
  & $\nu$       & 1.367 (0.323) & 1.842 & 77.8 \\
  & $t_0$       & 14.986 (0.332) & 1.341 & 94.9 \\
  & $c^{\ast}$  & 0.297 (0.022) & 0.072 & 73.3 \\
\hline
\multirow{4}{*}{Gamma (misspecified)}
  & $\Ro$       & 2.988 (0.026) & 0.092 & 86.9 \\
  & $\nu$       & 1.231 (0.208) & 0.751 & 78.7 \\
  & $t_0$       & 14.867 (0.329) & 1.269 & 96.2 \\
  & $c^{\ast}$  & 0.293 (0.017) & 0.057 & 80.3 \\
\hline
\multirow{4}{*}{Homogeneous (reference)}
  & $\Ro$       & 2.944 (0.020) & 0.073 & 19.5 \\
  & $\nu$       & --           & --   & -- \\
  & $t_0$       & 14.871 (0.320) & 1.169 & 92.5 \\
  & $c^{\ast}$  & 0.245 (0.002) & 0.008 &  0.0 \\
\hline
\end{tabular}
\end{table}

The two heterogeneous fits give similar median estimates of  $\hat{\Ro}$ $t_0$ and $c^{\ast}$, all within $2.5\%$ of the true values. They differ on $\nu$: the correct lognormal fit gives a median of $1.367$ (relative bias $-3.3\%$), the misspecified gamma fit gives $1.231$ ($-13.0\%$). Both heterogeneous fits have wide CIs for $\nu$ (width $0.75$ to $1.84$) and coverage between $77\%$ and $79\%$. No misspecification can be identified from a single epidemic by goodness-of-fit alone.

\begin{figure}[!h]
\centering
\includegraphics[width=0.95\linewidth]{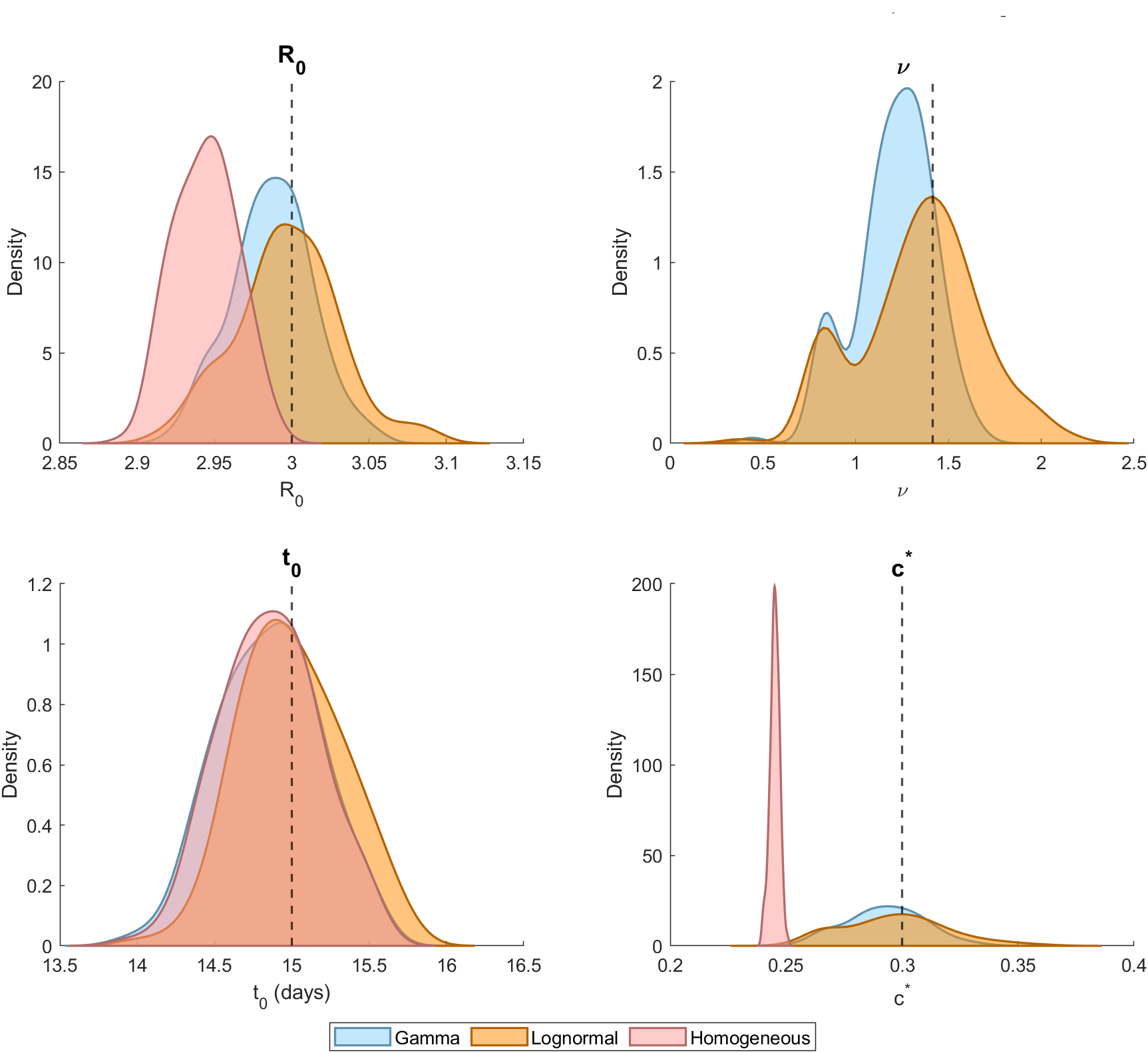}
\caption{
\textbf{Distributions of single-epidemic MLE estimates for lognormal-generated data ($\nu_{\text{true}} = 1.414$).}
Blue: gamma (misspecified). Orange: lognormal (correct). Pink: homogeneous (reference). Dashed vertical lines indicate the true values. The two heterogeneous densities overlap substantially in $\Ro$, $\nu$, $t_0$ and $c^{\ast}$. The homogeneous fit collapses ${c}^{\ast}$ to a tight peak at $0.245$.}
\label{fig:density_single_logn}
\end{figure}

The homogeneous fit behaves differently. It collapses ${c}^{\ast}$ to a tight distribution at $0.245$ (true value $0.30$, relative bias $-18\%$) with $0\%$ coverage, and biases ${\Ro}$ to $2.944$ ($-1.9\%$) with $19.5\%$ coverage. The heterogeneity signal is absorbed into the intervention parameter, recovering the misattribution mechanism reported in \cite{Mohammed2025}. Fig~\ref{fig:density_single_logn} shows the homogeneous ${c}^{\ast}$ density as a sharp peak well separated from both heterogeneous distributions.

\subsection{Single-epidemic forecasts}

\begin{figure}[!h]
\centering
\includegraphics[width=0.95\linewidth]{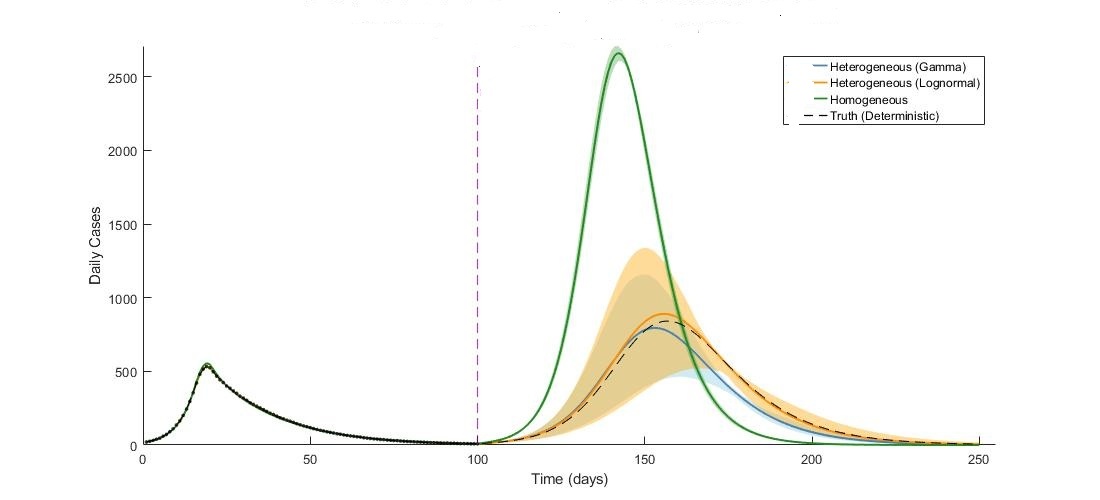}
\caption{
\textbf{Single-epidemic forecast trajectories for lognormal-generated data ($\nu_{\text{true}} = 1.414$, NPIs lifted at day~100).}
The black dashed line is the deterministic truth; shaded bands are 95\% confidence envelopes across replicated fits. The lognormal fit (orange) and the gamma fit (blue) both lie close to the truth in median trajectory and overlap it in confidence band. The homogeneous fit (green) produces a much higher and earlier peak.}
\label{fig:forecast_single_logn}
\end{figure}

To translate parameter estimates into forecasts, we lift NPIs at day~$T=100$
and simulate projected incidences to day~$250$ using each replicate's fitted parameters.
Fig~\ref{fig:forecast_single_logn} shows the resulting trajectories. The
two heterogeneous fits produce mean forecast trajectories that lie close to
the deterministic truth across the entire forecast window. Median peak height
differs from truth by $+2.9\%$ for the correct lognormal fit and by $-6.5\%$ for the misspecified gamma fit; median final size differs by $+1.7\%$ and $-12.5\%$ respectively (noting that the single-epidemic heterogeneous medians have wide interquartile ranges; see Supplementary Table~S8). The homogeneous fit overshoots the peak by $+216\%$ and the final size by $+88\%$, with no overlap between its 95\% confidence band and the truth across most of the forecast window. The misspecified gamma forecast error is therefore about $33$ times smaller in magnitude than the homogeneous error on peak height, and about $7$ times smaller on final size.

\subsection{Two-epidemic inference}\label{sec:two_misspec}

In Section~\ref{sec:single_misspec} we found that the misspecification of a lognormal distribution as gamma could not be diagnosed by fitting a single epidemic trajectory as the $95\%$ confidence intervals for estimated parameters were large enough to absorb any biases. Also, the $95\%$ confidence envelopes across replicated fits were wide enough to absorb the relatively small divergence in forecasts beyond the training time-window of 100 days. The wide confidence intervals are consistent with an extensive investigation conducted recently in \cite{Mohammed2025} for SEIR models with gamma-distributed susceptibility. The study addressed a positive correlation between the coefficient of variation estimates, $\hat{\nu}$ and the NPIs floor $ \hat{c^{\ast}}$. Due to this correlation, it is plausible that, in the current analysis, the optimiser is using a trade off between heterogeneity and intervention strength to minimise the overall impact that distribution misspecification exerts on epidemic trajectories. Following \cite{Mohammed2025}, we now extend our distribution misspecification analyses by conducting inference on pairs of epidemics with shared parameters as a strategy to alleviate the correlation and narrow confidence intervals. 

To apply the two-epidemic framework, we simulate pairs of epidemics with the same parameter values but different initial conditions (large seed: $\Ei_0 = 1000$, $\Ii_0 = 400$; small seed: $\Ei_0 = 100$, $\Ii_0 = 40$). We then estimate the same parameter vector $\boldsymbol{\theta}=(\Ro,\, \nu,\, t_0,\, c^{\ast})$ as defined in \eqref{eq:theta} by jointly fitting the two epidemics. The joint log-likelihood is the sum over both trajectories.

% \subsubsection{Identifiability gain under correct specification} 
Table~\ref{tab:two_logn_correct} compares single- and two-epidemic inference when the model is correctly specified. The CI width for $\nu$ shrinks from $1.842$ (single) to $0.070$ (two), a factor of $26$, and coverage rises from $77.8\%$ to $96.5\%$. A similar gain is seen for $c^{\ast}$ (width $0.072 \to 0.008$, coverage $73.3\% \to 97.0\%$). As in \cite{Mohammed2025}, joint two-epidemic inference resolves the issue with wide confidence intervals reported in the single-epidemic analysis when the distribution family is correctly specified.

\begin{table}[!ht]
\centering
\caption{
\textbf{Single vs.\ two-epidemic inference under correct specification (lognormal-generated data, lognormal fit).}
Median (SD), mean 95\% CI width, and empirical coverage across 200 replicates.}
\label{tab:two_logn_correct}
\begin{tabular}{|l|l|l|l|l|l|l|}
\hline
\multirow{2}{*}{\textbf{Parameter}} & \multicolumn{2}{|l|}{\textbf{Single epidemic}} & \multicolumn{2}{|l|}{\textbf{Two epidemics}} & \multicolumn{2}{|l|}{\textbf{Coverage (\%)}} \\
%\hline
\cline{2-7}
 & Median (SD) & Width & Median (SD) & Width & Single & Two \\
\hline
$\Ro$        & 2.996 (0.034) & 0.116 & 3.000 (0.010) & 0.041 & 84.7 & 96.0 \\
\hline
$\nu$        & 1.367 (0.323) & 1.842 & 1.414 (0.018) & 0.070 & 77.8 & 96.5 \\
\hline
$t_0$        & 14.986 (0.332) & 1.341 & 15.012 (0.196) & 0.693 & 94.9 & 92.0 \\
\hline
$c^{\ast}$   & 0.297 (0.022) & 0.072 & 0.300 (0.002) & 0.008 & 73.3 & 97.0 \\
\hline
\end{tabular}
\end{table}

% \subsubsection{Joint fitting exposes misspecification} 
Table~\ref{tab:two_logn_all} compares the three fits in the two-epidemic setting. The misspecified gamma fit retains narrow CIs (width $0.045$ for $\nu$) but the estimates are biased: $\hat{\nu} = 1.194$ versus the true $1.414$ (relative bias $-15.5\%$), and coverage for $\Ro$ and $\nu$ collapses to $0.0\%$. The single-epidemic compensation pathway has been removed; the structural mismatch is forced into a precisely biased point estimate. Joint fitting therefore serves a dual role: it sharpens inference under correct specification and it exposes misspecification under incorrect specification through coverage collapse.

\begin{table}[!ht]
\centering
\caption{
\textbf{Two-epidemic inference for lognormal-generated data, fitted with lognormal, gamma and homogeneous models.}
Joint fit across large-seed and small-seed epidemics. Convergence: 200/200 (heterogeneous fits), 195/200 (homogeneous).}
\label{tab:two_logn_all}
\begin{tabular}{|l|l|l|l|l|}
\hline
\textbf{Model} & \textbf{Parameter} & \textbf{Median (SD)} & \textbf{CI width} & \textbf{Coverage (\%)} \\
\hline
\multirow{4}{*}{Lognormal (correct)}
  & $\Ro$       & 3.000 (0.010) & 0.041 & 96.0 \\
  & $\nu$       & 1.414 (0.018) & 0.070 & 96.5 \\
  & $t_0$       & 15.012 (0.196) & 0.693 & 92.0 \\
  & $c^{\ast}$  & 0.300 (0.002) & 0.008 & 97.0 \\
\hline
\multirow{4}{*}{Gamma (misspecified)}
  & $\Ro$       & 2.944 (0.010) & 0.037 &  0.0 \\
  & $\nu$       & 1.194 (0.012) & 0.045 &  0.0 \\
  & $t_0$       & 15.154 (0.203) & 0.689 & 82.0 \\
  & $c^{\ast}$  & 0.297 (0.002) & 0.008 & 66.0 \\
\hline
\multirow{4}{*}{Homogeneous (reference)}
  & $\Ro$       & 3.010 (0.022) & 0.029 & 45.9 \\
  & $\nu$       & --           & --   & -- \\
  & $t_0$       & 10.051 (0.310) & 0.312 &  0.0 \\
  & $c^{\ast}$  & 0.236 (0.002) & 0.004 &  0.0 \\
\hline
\end{tabular}
\end{table}

\begin{figure}[!h]
\centering
\includegraphics[width=0.95\linewidth]{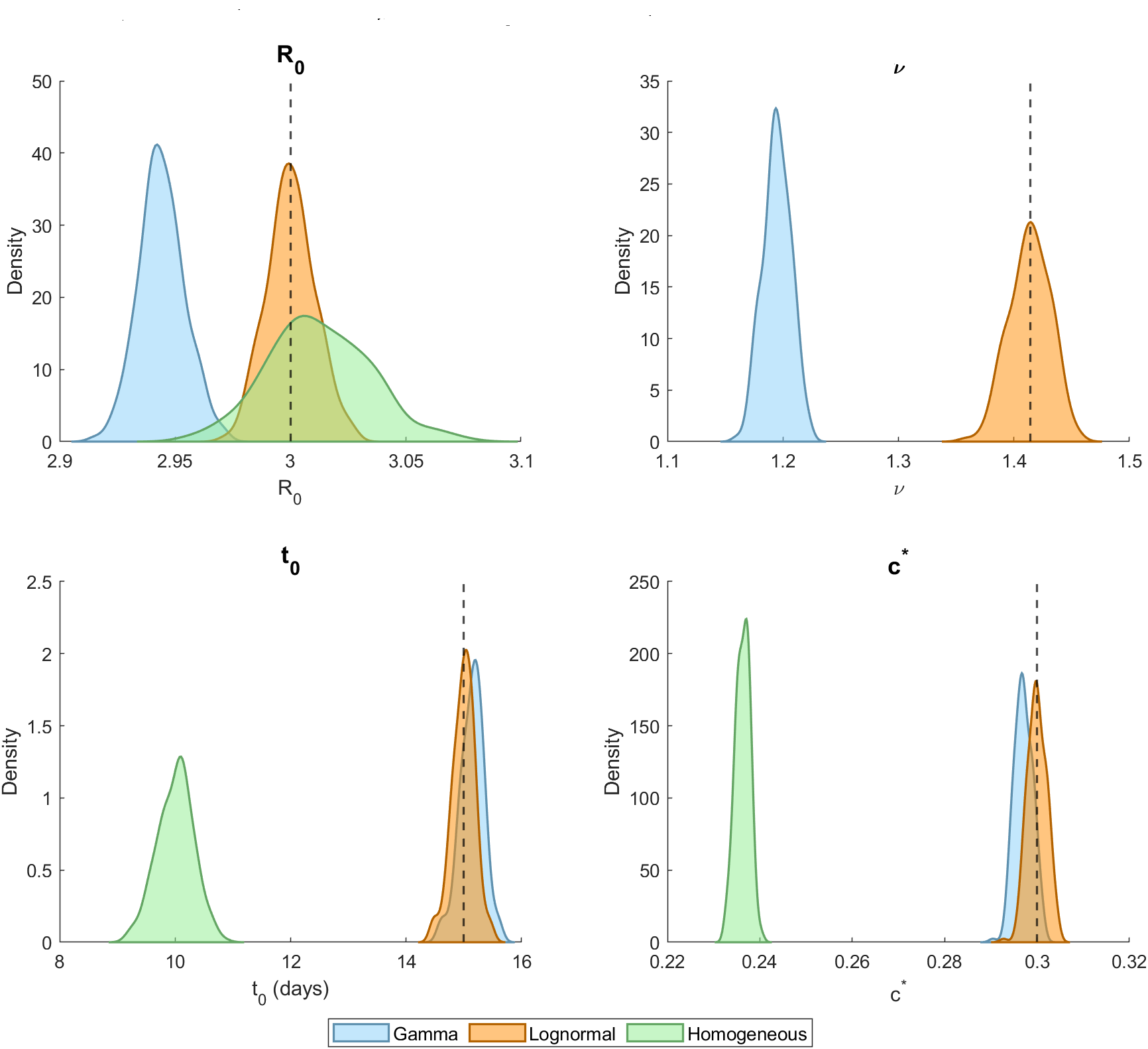}
\caption{
\textbf{Distributions of two-epidemic MLE estimates for lognormal-generated data ($\nu_{\text{true}} = 1.414$).}
Blue: gamma (misspecified). Orange: lognormal (correct). Green: homogeneous (reference). Dashed vertical lines indicate the true values. The misspecified gamma fit collapses to a sharp peak at $\hat{\nu} \approx 1.19$, well separated from the correct lognormal density. The homogeneous fit fails on $t_0$ and $c^{\ast}$.}
\label{fig:density_two_logn}
\end{figure}

The homogeneous fit fails on three of the four parameters in the joint setting: $\hat{{t}}_0 = 10.05$ days (true 15, relative bias $-33\%$, $0\%$ coverage), ${\hat{c}}^{\ast} = 0.236$ ($-21\%$, $0\%$ coverage), and ${\hat{\Ro}} = 3.010$ with only $46\%$ coverage. Fig~\ref{fig:density_two_logn} shows the corresponding MLE densities.

% \subsection{Directionality of misspecification bias}

The two-epidemic experiment reveals a direction in the misspecification bias on $\nu$. When lognormal data are fitted with a gamma model the median ${\nu}$ is pulled downward to $1.194$ (relative bias $-15.5\%$). The gamma family lacks the heavy upper tail that drives lognormal-truth dynamics, and the optimiser compensates by lowering inferred heterogeneity. The mismatch is concentrated in $\nu$: $\Ro$ shifts by $-1.9\%$ and $c^{\ast}$ by $-1.0\%$ in the two-epidemic setting (Table~\ref{tab:two_logn_all}).

% The gamma-truth experiment shows the opposite signature (\nameref{S1_Appendix}, Supplementary Table~S7): $\hat{\nu}$ is pushed upward to $1.781$ (relative bias $+25.9\%$). The lognormal's heavier-than-gamma tail implies stronger depletion than the data support, and the optimiser inflates $\nu$ to compensate. The two directions of misspecification therefore produce structural and concentrated bias on $\nu$, but the magnitude is asymmetric: $|{+25.9\%}| / |{-15.5\%}| = 1.7$, so fitting a heavier-tailed family to lighter-tailed data produces $1.7\times$ the relative shift in $\hat{\nu}$ compared with the converse direction.

\subsection{Two-epidemic forecasts}\label{sec:forecasts}

\begin{figure}[!h]
\centering
\includegraphics[width=\textwidth]{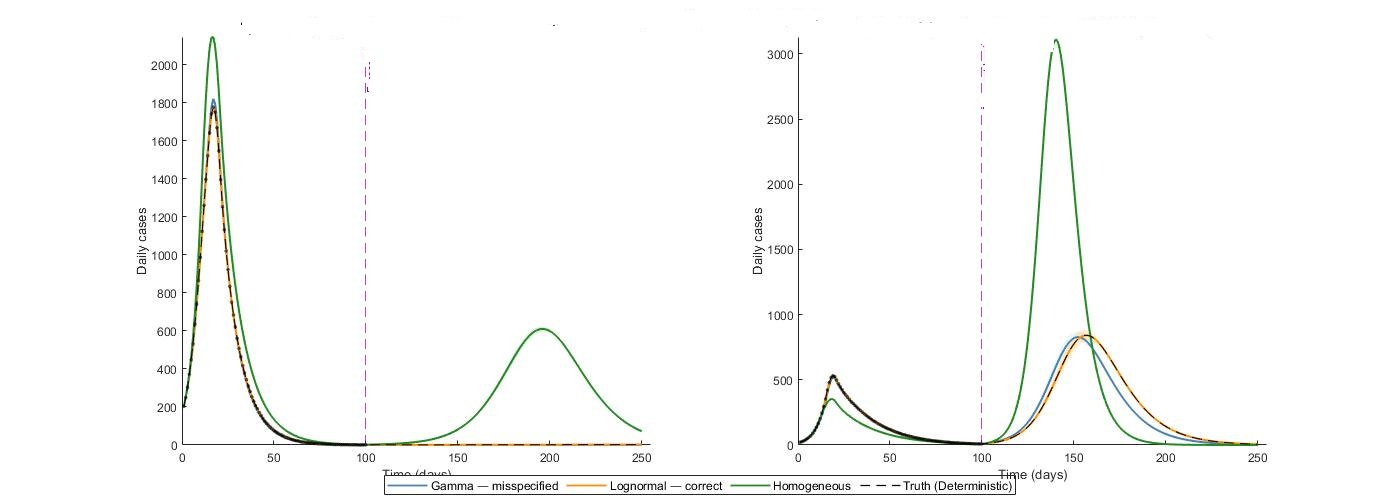}
\caption{
\textbf{Two-epidemic trajectory fits and forecasts for lognormal-generated data ($\nu_{\text{true}} = 1.414$, NPIs lifted at day~100 in each epidemic).}
The black dashed line is the deterministic truth; shaded bands are 95\% confidence envelopes across replicated fits. The correctly specified lognormal fit (orange) tracks both waves accurately. The misspecified gamma fit (blue) tracks the truth closely in both epidemics in the 100 day fitted training window but the forecast shows an undershoot by the gamma misspecified model. The homogeneous fit (green) already departs from the truth during the training window and overshoots substantially in both epidemics, with the large-seed epidemic showing the largest relative error ($+25{,}881\%$ peak height vs.\ $+270\%$ in Epidemic~2).}
\label{fig:2epi_forecast_logn}
\end{figure}

To translate parameter biases into forecasts, we extend each two-epidemic simulation by lifting NPIs from day 100 onwards and projecting to day 250. Forecasts are computed per replicate using its joint-fit MLEs. Fig~\ref{fig:2epi_forecast_logn} shows the resulting trajectories.

The forecast errors at $\nu = 1.414$ differ sharply between the two epidemics. In the small-seed epidemic, all three fits behave similarly to the single-epidemic case: the correct lognormal fit median peak-height error is $+0.3\%$, the misspecified gamma fit is $-1.3\%$, and the homogeneous fit is $+270.3\%$. In the large-seed epidemic the result is different. Susceptible depletion during the
training window is nearly enough to prevent a second wave whose truth peak height is only $2.4$~cases\,day$^{-1}$; small biases in ${\nu}$ therefore propagate into large relative errors on this small denominator.
The correct lognormal fit gives a median peak-height error of $+1.5\%$ in the large-seed epidemic, but the misspecified gamma fit gives $-71.9\%$, and the homogeneous fit gives $+25881\%$. The gamma-misspecification cost appears therefore not negligible in the large-seed epidemic even though it is small in the small-seed case, but the large relative magnitude partly reflects the small truth denominator rather than a proportionately large absolute forecast miss.

A consistent ordering nonetheless emerges across both epidemics and all four CV values in the sensitivity sweep (Supplementary Tables~S8 and~S9). Defining the absolute median peak-height error as the metric, the homogeneous error exceeds the misspecified gamma error by factors ranging from $4.4\times$ ($\nu = 0.5$ in the single-epidemic analyses) to $2500\times$ ($\nu = 2$ in the large-seed epidemic of two-epidemic analyses). The error magnitude ordering (homogeneous $>$ misspecified heterogeneous $>$ correctly specified heterogeneous) holds in every case.

\subsection{Asymmetry across gamma-lognormal scenarios}

Numerical experiments conducted with gamma-truth show the reverse signatures to the lognormal-truth results reported above, but biases are more severe (\nameref{S1_Appendix}). In two-epidemic inferences with $\nu=1.414$, for instance, when synthetic data generated with gamma-distributed susceptibility are fitted by a lognormal model, the estimated ${\nu}$ is pushed upward to $1.781$ (relative bias $+25.9\%$) (Supplementary Table~S7), while the reverse analysis gave a relative bias of $-15.5\%$ (Table~\ref{tab:two_logn_all}). The lognormal's heavier tail implies greater susceptibility depletion than the synthetic data support, and the optimiser inflates $\nu$ to compensate. 

Two-epidemic forecasts with gamma-truth (Supplementary Fig~S15) show the same ordering of qualities as in lognormal-truth, but with a larger distribution misspecification signal. Again in the case $\nu=1.414$, fitting the lognormal model to gamma-generated data overshoots the small-seed epidemic
median peak-height by $+5.7\%$, compared with a $-1.3\%$ undershoot in the converse case at the same $\nu$. Given the more severe biases that result from fitting the lognormal model, it appears commendable to adopt gamma distributions as default for modelling epidemics where the susceptibility distribution is unknown.

\section{Discussion}\label{sec:discussion}

This computational study assessed the impacts of susceptibility distribution
misspecifications on the inference of epidemic model parameters and epidemic forecasting. We implemented SEIR models with gamma-distributed and lognormal-distributed susceptibility, as well as the classic homogeneous version. We then generated synthetic datasets with both the gamma and lognormal models, and performed inference and forecasts under correct and misspecified susceptibility distributions. We provided a comprehensive assessment of the results. 

% \subsection{The tail matters} 
For the same mean and coefficient of variation, the lognormal susceptibility distribution produces a larger and later epidemic than the gamma, with the difference driven by the heavier right tail of the lognormal (Fig~\ref{fig:moments_comparison}). At $\nu = 1.414$ and $\Ro = 3$, the lognormal peak is $16.1\%$ higher than the gamma and the final size is approximately $30\%$ larger (Fig~\ref{fig:trajectory}). This is consistent with the characterisation of gamma as an eigen-distribution that provides maximal braking under selective depletion \cite{Rose2021HeterogeneityModels}.

% \subsection{Single-epidemic inference is non-identifiable for heterogeneous fits} 
In a recent paper \cite{Mohammed2025}, we noted that inferring the susceptibility distribution from a single epidemic dataset is challenging due to various parameter correlations. Of special interest was a positive correlation between the coefficient of variation ($\nu$) and the proportion of contacts that remain active during a non-pharmaceutical intervention such as lockdown ($c^{\ast}$). This correlation supplies a compensation pathway that lets the optimiser absorb structural mismatches (such as the distribution misspecifications considered here) by trading off heterogeneity against intervention strength. As a result, both heterogeneous fits at $\nu = 1.414$ produce wide CIs for $\nu$ (width $0.75$ to $1.84$, Table~\ref{tab:single_logn}) and coverage between $77\%$ and $79\%$. The misspecified gamma fit shows a shift in ${\nu}$ (median $1.231$ versus correct-fit $1.367$) but recovers $\Ro$, $t_0$ and $c^{\ast}$ within $1\%$ of the true values. Standard goodness-of-fit diagnostics cannot distinguish the two heterogeneous fits from a single epidemic.

% \subsection{Multi-epidemic inference resolves identifiability and exposes misspecification}
Joint fitting across two concurrent epidemics with shared parameters narrows the CI for $\nu$ by a factor of $26$ and raises coverage to $96.5\%$ when the model is correctly specified (Table~\ref{tab:two_logn_correct}). When the gamma model is fitted to lognormal-generated data, the joint constraint forces the estimator into a precisely biased point: ${ \hat{\nu}} = 1.194$ (relative bias $-15.5\%$, coverage $0\%$), with biases in $\Ro$ and $c^{\ast}$ of $-1.9\%$ and $-1.0\%$ respectively (Table~\ref{tab:two_logn_all}). The compensation pathway available in the single-epidemic setting is removed, and the structural mismatch is concentrated in ${\nu}$. Joint fitting therefore plays a dual role: it sharpens identifiability when the model class is correct, and it exposes misspecification through coverage collapse on $\nu$ and $\Ro$ when the model class is wrong.

% \subsection{Highlight with Mohammed et al.\ work}
In \cite{Mohammed2025} we demonstrated that fitting heterogeneous models to homogeneous-truth data recovers the truth, while fitting homogeneous models to heterogeneous-truth data produces strongly biased estimates and poor forecasts. The present results refines the analysis to
misspecification within the heterogeneous class. Fitting the wrong distribution family can bias the inferred ${\nu}$, but this is not typically detectable unless multiple epidemics with shared parameters are fitted concurrently. The forecast implications of the susceptibility distribution misspecification encountered here are nevertheless substantially smaller than the cost of omitting heterogeneity all together.

% \subsection{Asymmetry across misspecification directions}
The gamma-truth experiment (\nameref{S1_Appendix}) shows the opposite directional signature on ${\nu}$, with magnitude $1.7\times$ larger ($+25.9\%$ versus $-15.5\%$). The asymmetry reflects the heavier right tail of lognormal distributions: when the lognormal model is fitted to synthetic data generated with lighter-tailed gamma, the optimiser inflates $\nu$ to compensate for excess susceptibility depletion, more strongly than the converse
compensation. At the forecast level, fitting a heavier-tailed family to
lighter-tailed data overshoots the truth epidemic peak height; fitting a
lighter-tailed family to heavier-tailed data undershoots peak height. Comparing the two directions, errors are smaller when gamma distributed susceptibility is used to fit the synthetic data. On this basis, we cautiously recommend the gamma family as default for modelling epidemics where the susceptibility distribution is unknown. This adds to previously exposed  that make gamma the distributions of choice for modelling susceptibility, such as the eigen-distribution property \cite{Rose2021HeterogeneityModels} and the computational tractability \cite{Novozhilov2008OnPopulation, Antonio2022Reinfection}.

The numerical experiments compared gamma and lognormal specifications only. Bounded distributions such as the beta, and heavy-tailed distributions with infinite variance such as the positive stable, would require modifications to the moment-matching strategy and remain for future work. All experiments used synthetic data with known truth; application to real multi-wave epidemic data is the natural next step. The two-epidemic design assumes parameters are exactly shared across epidemics, which may not hold in practice \cite{ibrahimThesis}; exploring partial sharing is also a useful extension.

\section*{Supporting information}
% \paragraph*{Supporting information.}
\label{S1_Appendix}
% \textbf{Supplementary material.} 
Contains the full derivation of the distribution-agnostic discretisation framework (Sections~S1--S3), existence and uniqueness proofs for the log-affine transformation, distribution-specific ingredients for both gamma and lognormal families (Sections~S4--S5), numerical illustrations of the discretisation (Section~S6), distribution-level validation (Section~S7, Table~S1), trajectory validation results (Section~S8, Tables~S2--S4, Fig~S10), additional inter-family trajectory comparison (Section~S9, Fig~S11), gamma-truth inference results including single-epidemic (Section~S10, Table~S5, Fig~S12--S13) and two-epidemic (Tables~S6--S7, Figs~S14--S15), and sensitivity sweeps across heterogeneity levels (Section~S11, Tables~S8--S9).

\section*{Acknowledgments}

M.G.M.G. is partially funded by FCT -- Funda\c{c}\~{a}o para a Ci\^{e}ncia e a Tecnologia, I.P., projects UIDB/00297/2020 and UIDP/00297/2020 (Center for Mathematics and Applications).
I.M. is funded by the Petroleum Technology Development Fund (PTDF), Nigeria.

\nolinenumbers

\bibliography{reference}

\end{document}